\documentclass{sigchi}

\setcopyright{none}

\usepackage{balance}       
\usepackage{graphics}      
\usepackage[T1]{fontenc}   
\usepackage{txfonts}
\usepackage{mathptmx}
\usepackage[pdflang={en-US},pdftex]{hyperref}
\usepackage{color}
\usepackage{booktabs}
\usepackage{textcomp}
\usepackage{subcaption}
\usepackage{microtype}        
\usepackage{ccicons}          

\def\plaintitle{Challenges of Bridging the Gap between Mass People and  Welfare Organizations in Bangladesh}
\def\plainkeywords{Street children, Social Welfare Organizations, Ergonomics, Human Computer Interaction, Sustainable Development.}

\makeatletter
\def\url@leostyle{%
  \@ifundefined{selectfont}{
    \def\UrlFont{\sf}
  }{
    \def\UrlFont{\small\bf\ttfamily}
  }}
\makeatother
\def\pprw{8.5in}
\def\pprh{11in}

\definecolor{linkColor}{RGB}{6,125,233}
\hypersetup{%
  pdftitle={\plaintitle},
  pdfkeywords={\plainkeywords},
  pdfdisplaydoctitle=true, 
  bookmarksnumbered,
  pdfstartview={FitH},
  colorlinks,
  citecolor=black,
  filecolor=black,
  linkcolor=black,
  urlcolor=linkColor,
  breaklinks=true,
  hypertexnames=false
}

\makeatletter
\def\@copyrightspace{\relax}
\makeatother
\begin{document}

\title{\plaintitle}

\numberofauthors{2}
\author{%
  \alignauthor{Alvi Md Ishmam\\
    \affaddr{Bangladesh University of Engineering and Technology}\\
    \affaddr{Dhaka, Bangladesh}\\
    \email{1305092.am@ugrad.cse.buet.ac.bd}}\\
  \alignauthor{Md Raihan Mia\\
    \affaddr{Bangladesh University of Engineering and Technology}\\
    \affaddr{Dhaka, Bangladesh}\\
    \email{htraihan@gmail.com}}\\
}

\maketitle

\begin{abstract}
Computing for the development of marginalized communities is a
big deal of challenges for researchers. Different social organizations
are working to develop the conditions of a specialized marginalized
community namely ’Street Children’, one of the most underprivileged
communities in Bangladesh. However, lack of proper engagement
among different social welfare organizations, donors, and
mass community limits the goal of development of street children.
Developing a virtual organization hub can eliminate the communication
gap as well as information gap by involving people of all
communities. However, some human imposed stigmas may often
limit the rate of success of potential virtual computing solutions
intended for organizations working with the marginalized communities,
which we also face in our case. After a partial successful
deployment, the design itself needs to be self comprehensive and
trustworthy in order to overcome the stigmas that demand a reasonable
amount of time. Moreover, after a wide scalable deployment, it
is yet to be investigated whether the design of our computational
solution can attain the goal for the facilitation of the organizations
so that those organizations can become more effective for the
development of street children than before.
\end{abstract}

\ccsdesc[300]{Human Centered Computing~Collaborative and Social Computing Design and Evaluation Method}
\ccsdesc[300]{Human-centered computing~Interaction Design}

\keywords{\plainkeywords}

\printccsdesc

\section{Introduction}

\textit {Sustainable development} \cite{sustainable} (such as economic, social, educational, etc.) is a major challenge all over the world. Sustainable development for society cannot be achieved when resource allocation consistently favors privileged communities only, while underprivileged communities are made to fend for themselves \cite{ict4d}. Experts across the world are now trying to eradicate such barriers against sustainable development via computing technologies \cite{effectiveComputing}. In doing so, they face unique human and socio-economic challenges \cite{humanfactors} that limit the potential of computing technologies from development to deployment phases. Such challenges are more prominent when target populations have special needs, for example, underprivileged children, in poorer countries.

Let us look at Bangladesh, which has $164$ million people and a density of $1,250$ people per square kilometer \cite{population}.  People from all over the country converge to Dhaka (the capital city) for livelihood, and most of them face severe economic hardships and exploitation. The most vulnerable part of people here are \textit{Street Children} (known as \textit{`tokai'} in Bengali), who suffer from starvation, physical abuse, sexual harassment, bonded labor, lack of health care and more. They often end-up committing crimes that warrant arrests, which further their spiral towards destruction.

Naturally, in a country like Bangladesh, or in any similar country with children still living on streets, sustainable development for them will only be a dream with the status quo. Furthermore, the poorer a country is, the more are the challenges and ensuing inaction from governments. While larger organizations such as  \textbf{UNICEF, JAAGO, Save the Children}, etc., are doing their part in Bangladesh, these alone are insufficient, and and a result, several Non-Governmental Organizations (called NGOs) mainly operated by young and educated people are now emerging and trying to tackle 
this problem, either independently by themselves, or by being a bridge between needy children and more established organizations \cite{brown1991bridging}. 

Our motivation in this paper is to analyze the operational mechanisms of these smaller NGOs that work with street children in Bangladesh and use findings to enhance their effectiveness. Despite common goals and objectives among multiple such NGOs, we find that there are subtle, and yet, tangible diversities among them, that lead to barriers in their effectiveness. Most critically, we found that there is always some form of a ``communication/ engagement gap" between NGOs, donors, and the general public. Unless there is sufficient engagement, credibility, accountability, and clearly defined agendas, donors are unwilling to financially support NGOs, without which NGOs cannot create impact for street children.

In this paper, we design, deploy, and evaluate a  digital and integrated organizational hub for a) addressing current deficiencies of small--scale NGOs working for street children in Bangladesh; and b) assisting donors and the general public stay engaged with the NGOs. Based on the study, our contributions are as follows:

 \begin{itemize}
 \item After several formal and informal discussions with correspondents from  NGOs, social workers, and street children, and based on membership sizes and operating structures, we first classified small-scale NGOs as micro and semi-micro organizations. Micro organizations are smaller in size and with limited financial resources, with about $20$ volunteers. To establish an initial foothold in the field, organizations prefer this model, and it is also easier for external entities (e.g., the authors of this paper) to engage with these NGOs. Semi-structured NGOs are larger with about $80$ volunteers, with more financial clout. In such NGOs, we identified complex rules, rigid traditions/ regulations, and significant diverse visions among members that often create unification challenges. Furthermore, convincing all members to adopt any new technology (e.g., our system) was much harder in this case. Designing technologies to accommodate both models is challenging and are elaborated later in the paper.
 
 \item We conducted two rounds of rigorous and semi-structured face-to-face interviews with three NGOs (two micro and one semi-micro) working for street children. The first interview helped us understand operating structures, challenges, needs, policies, working procedures, goals, traditions, values, and visions of NGO personnel. Analyzing these outcomes led us to the design of a unique and novel participatory social interactive platform specially designed for NGOs working towards the upliftment of street children. Our system enables NGOs to engage with local communities such as donors, the general public, interested volunteers, etc., by showcasing organizational activities and events. On the other hand, the general public can be a part of organizations through voluntary deeds and funding. After testing our system for $30$ days, each of the three NGOs was contacted for a second round of interviews for soliciting feedback, which as we report later in the paper is not only very encouraging in terms of donation and donor behavior, but also helped us identify unique ergonomic challenges.
 
\item We also incorporated two new organizations (that we did not interview before) - one working for street children similar to three organization and another working for orphans (i.e., an actual orphanage home). Both organizations actively used our system. More encouragingly, the orphanage received donations from new donors after it was being introduced via our system. The donations exhibit a degree of temporal alignment with the Arabic month of Ramadan (a holy month for Muslims), 


\item Finally, we provide a comprehensive discussion on all lessons learned so that ideas using computing technologies such as the ones we develop could be scaled and hopefully find wider acceptance, for the upliftment of vulnerable communities like children living on streets in low-income countries. Based on our fieldwork,  we have added some valuable insights in light of existing models on ICT adoption in non--profit organizations that might be valuable for community stakeholders, designers, social entrepreneurs, researchers, and practitioners.
 \end{itemize}


\section{Related Work}
\label{sec:relat}

The HCI community has a significant interest in developing computing solutions for the enhancement of marginalized communities such as refugees, women, religious minorities, etc. However, the needs of each community are distinct, and also economic and cultural sensitivities do play a role in any solutions design to offer help. In this paper \cite{DIKC}, some ICT enabled institutions named Digital Knowledge Information Center (DKIC) are used in the context of Latin America, (Argentina) to improve literacy, e-literacy and provide useful skills for the street children. The DKIC focuses on the facilities that is required to fulfil the strategies taken for street children. This information center is operated by the government agencies, while in our solution we focus on the struggling NGOs with a goal of effective engagements of the stakeholders to alleviate financial and organizational challenges in a 24-7 interactive platform.


To improve the educational experience of the marginalized children of rural India, an interactive teaching learning tool with multimedia applications are introduced here \cite{interactivetool}. BingBee, \cite{slay2006bingbee} is an information kiosk, deployed among the street children of South Africa to improve the educational experiences. In a similar context, an ethnographic study of the homeless people of Los Angeles reveals how the technology is owned and used by homeless people by enabling social ties \cite{homelessness}. A qualitative study \cite{perception} on homeless people in Los Angeles details how technologies can improve their lives. The study specifically investigated how homeless people use computing devices and mobile phones (which is not uncommon for them to have) for earning money, searching for jobs, socializing and even sleeping at night. However, it is important to point out that such digital devices are simply unaffordable for homeless people in poorer countries, and more so unaffordable for street children, and even for volunteers that want to help them due to low--income settings.

In western countries, the structure of non--profit organizations and fundraising processes are different from those in third world countries. In these studies, \cite{goecks2008charitable,merkel2007managing} how participatory design can be used to empower organizational activities are discussed. We see, upon surveying related work that there are very limited studies, on how computing technologies can assist welfare organizations. This may be because in Western and Far Eastern societies, welfare organizations are relatively richer and they can afford to pay for customized technologies by themselves. With more taxpayer money, governments also can chip in with resources. In the paper \cite{Organization}, two non-profit voluntary organizations dedicated to homeless people in a US metropolitan city are studied. The findings are related to challenges emanating from the adoption of technologies, how the nature of volunteerism impacts these technologies, and how governmental assistance interacts with services offered by welfare organizations.

From the perspective of improving the lives of children, there are some very interesting research studies are going on. For example, the government of Cambodia \cite{combodia} has taken an initiative of counting homeless adolescent aged between $13$ to $17$ across seven cities of the country using manual counting and forecasting.  In Germany, there is a project that connects children and adult computer club participants in a participatory online platform called ``come\_NET''  \cite{comenet} to promote sharing of ideas across borders, supporting collaboration, and enhancing life skills in local neighborhoods.  A five-week-long qualitative study on children in a refugee camp in Palestine used 3D modeling and printing technologies for education \cite{palestine}. These types of technologies focus only on the specific need on the focus group with customized solution. However, we propose a theory driven unique model for connecting the all stakeholders of society for sustainability that can be applied to similar context (e.g., orphan).

While the technologies surveyed above (for refugees, homeless people, and children) are all exciting ideas, the problem addressed in our paper is unique. Here, as focused in Bangladesh, we consider a real case of NGOs having a very low number of volunteers, funds, and visibility with almost no governmental assistance. Providing food, clothing, shelter, and medicines for street children are major challenges for these NGOs. Furthermore, the plight of such children and NGOs in poorer countries has not been formally studied yet. As such, the interviews conducted in this paper, challenges of NGOs in this space, trends reported after interviews, our system design, and results on using our systems by the NGOs -- are all very unique, and can contribute to peer researchers attempting to address similar problems in poorer countries.

\section{Welfare Organizations We Engaged with and Our Context}
\label{sec:Related Context}
It is sadly difficult to determine the actual number of street children in Bangladesh, though a report \cite{Street} indicates that it was about $1.5$ million in $2015$. However, with more than a few organizations (a few big, and some small) working for the welfare of street children, we wondered why are there so many still on the streets begging for food, impoverished and vulnerable to exploitation in Dhaka? 

\begin{figure}[!h]

\minipage{0.32\linewidth}
  \includegraphics[width=\linewidth, height = 2.5cm]{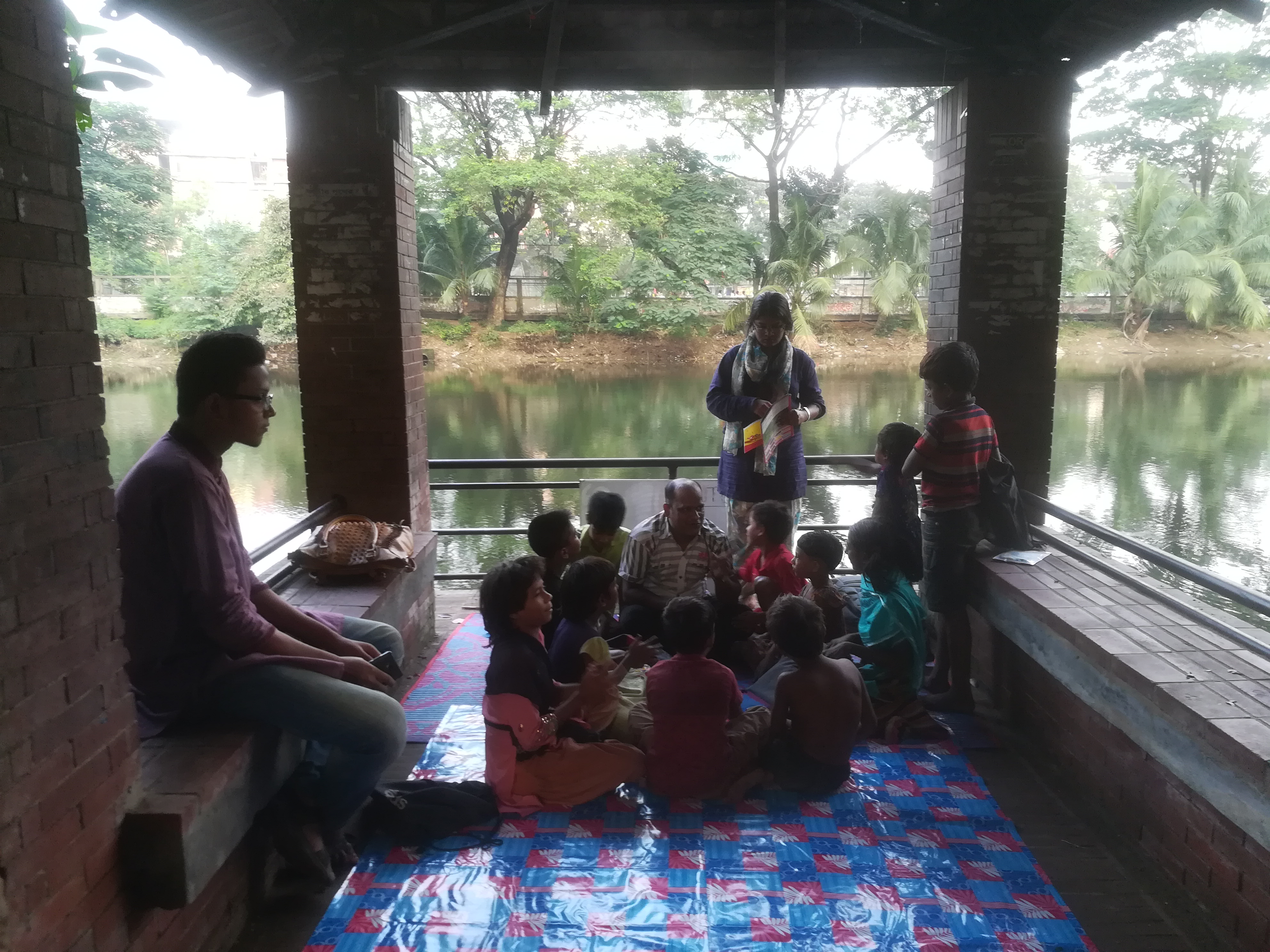}
  \subcaption{A schooling session arranged by an organization named ``Putul''}\label{fig:A schooling session arranged by an organization}
\endminipage\hfill
\minipage{0.32\linewidth}
  \includegraphics[width=\linewidth, height = 2.5cm ]{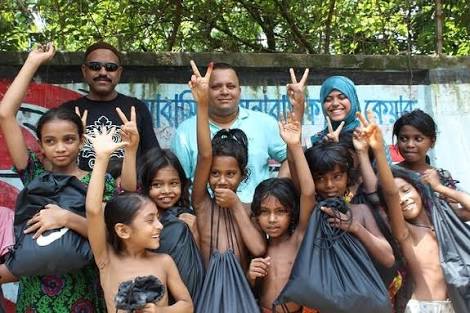}
  \subcaption{A campaign arranged by an organization named ``Child Care''}\label{fig:A schooling session arranged by an organization project khata kolom}
\endminipage\hfill
\minipage{0.32\linewidth}%
  \includegraphics[width=\linewidth,  height = 2.5cm]{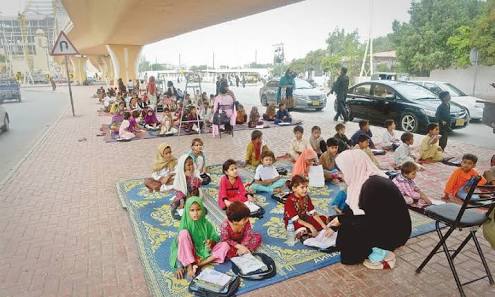}
  \subcaption{Free schooling arranged by an organization named ``Shishu Bikash''}\label{fig:Free schooling arranged by an organization named ``Shishu Bikash''}
\endminipage
\caption {Schooling and campaign arranged by different voluntary organizations}
\label{fig:Schooling and campaign arranged by different voluntary organizations}
\end{figure}

\begin{figure*}[h!]
     \centering
      \includegraphics[width = \textwidth]{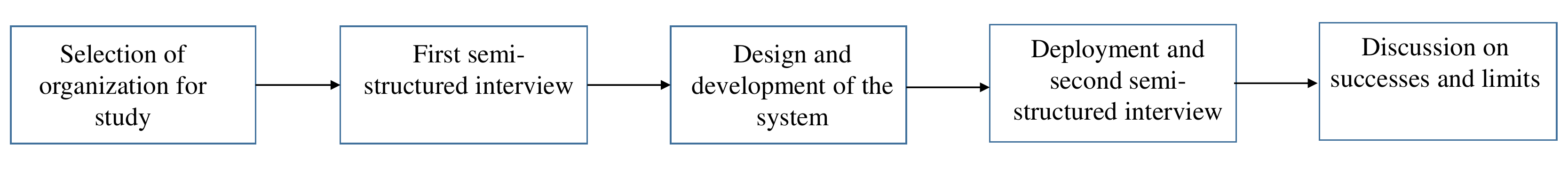}
    \caption{Workflow of our study}
    \label{fig:Workflow of This Study}
\end{figure*}

To answer this question, we chalked out a process to primarily interact with NGOs (Non-Governmental Organizations) themselves, along with street children, since any focus only on the latter group may help only identify problems, but not solutions. However, instead of blindly hunting down NGOs, we decided to narrow our search to ones run by younger people -- like recent university grads, or those still studying at universities or freelance voluntary services. It is a fact that these NGOs are critically important for improving lives of street children, even though they experience more financial, time, and other resource constraints compared to more established organizations (e.g., the UNICEF), therefore, we forecasted that these NGOs may be willing to give us time for our study. Furthermore, it is natural to infer that these smaller-scale NGOs will also be more amenable to experimenting with our technologies, and adapt their existing ideas and execution mechanisms accordingly, for the greater good.

The first thing we found out that there was no established database anywhere on NGOs working for street children in Bangladesh. We also found that common citizens across all age groups and wealth status simply had no idea about these NGOs, or how to reach them, beyond a basic notion that ``{\em such organizations exist somewhere in Dhaka and are doing something}". This was sad for us to discover. We then had to query our friends, and also some street children who knew NGO members. We also visited many universities with social science programs to talk to faculty and students there. After five weeks of effort, we found three NGOs whose members were willing to engage with us.

These three organizations in our study are \textit{Putul} (meaning Doll), \textit{Project Child Care} (formerly called {\em Khata Kolom}, meaning  \textit{Book and Pen}), and \textit{Shishu Bikash} (meaning Blooming Child). Fig. 
\ref{fig:A schooling session arranged by an organization}, \ref{fig:A schooling session arranged by an organization project khata kolom}, and \ref{fig:Free schooling arranged by an organization named ``Shishu Bikash''} depict different sessions organized by these three NGOs. Note that all images were captured with the approval of NGO members, and they permitted us to use these images for research purposes. We point out that \textit{Putul} and \textit{Project Child Care} have around $20$ volunteering members, which we classify as micro-organizations. These are relatively new NGOs and are in operation for less than three years. \textit{Shishu Bikash} is a more established    NGO with a sound operating structure for several years and has more than $80$ volunteers. We classified this as a semi-micro organization.

\begin{figure}[h!]
             \centering
             \vspace{1ex}%
   \includegraphics[width = \linewidth]{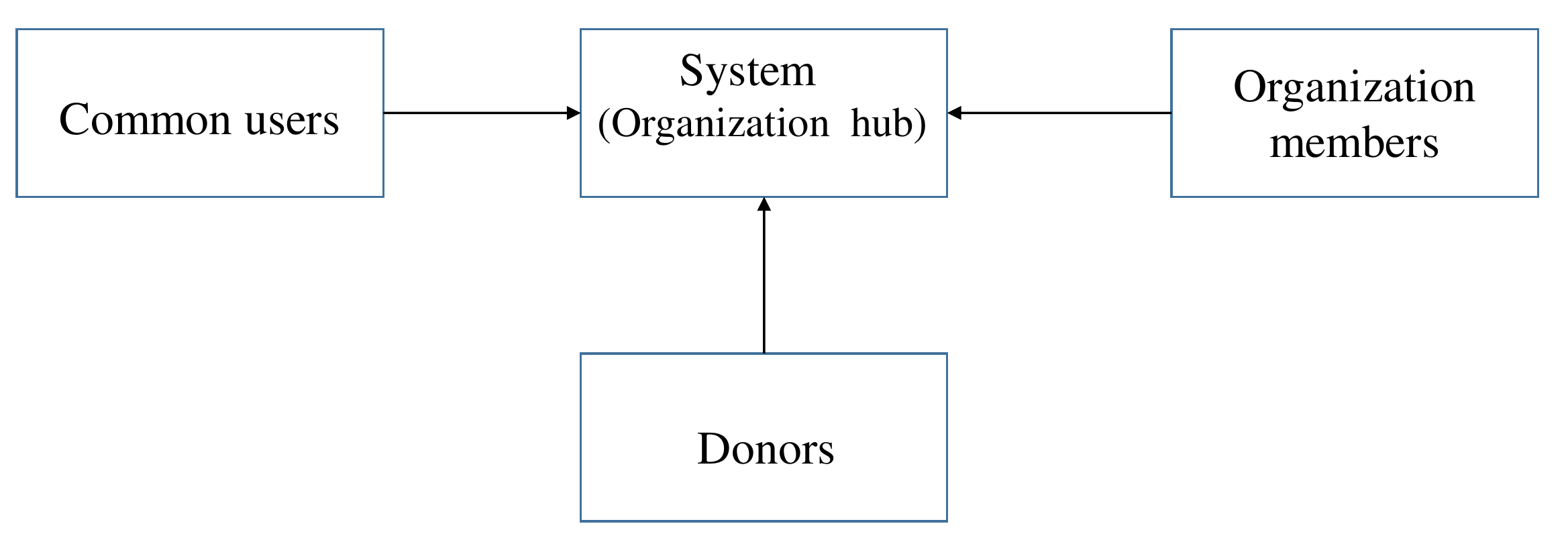}         
            \caption{Block diagram of our proposed system}
            \label{fig:Block Diagram}
             \end{figure}
             
We found out that each organization caters to a specific geographic zone in Dhaka, and they serve anywhere from $50$ to $100$ children living on the streets. Activities include informal schooling sessions at specific times of the week, supplying medicines, providing clothes, feeding them nutritious meals, conducting special events on national holidays and other basic services. The organizations primarily depend on personal funds from volunteers and funds from donors to support these efforts. Naturally, the latter portion of funds is most vital for sustenance.

Anyone can become a volunteer once he/she agrees to the basic principles of the respective NGOs. Membership can be generally divided into three parts: advisory, volunteers, and administrative panel depending on the size. Despite having similar goals, there are mild differences in the organizational structure, meeting times, meeting types, degree of communication or cooperation among members, etc., that will be discussed later in the paper.

\section{Steps Executed in Our Research}
\label{sec:Steps}

In Fig \ref{fig:Workflow of This Study}, we present the workflow of our study. At the outset, we have conducted two rounds of semi-structured face-to-face interviews with personnel from these three organizations. In the first face-to-face interviews, we created a questionnaire with $15$ questions to get an overview of a particular organization's nature, working procedure, operating structure, goals, vision, policy, etc. We have also gathered basic metadata of the organizations such as the total number of members, the total number of children served, etc.

Then, based on these interviews and our analysis of responses, we design and deploy a digital hub to benefit these NGOs, and enabled them to access our system for $30$ days. Subsequently, we conduct a second round of interviews with the members of NGOs to understand their perspectives that lead us to more rigorous insights on how computing technologies can aid their operation. Related discussions and findings are elaborated later in the paper.

\subsection{First Round of Face-to-Face Interviews and Its Findings}

We visited each NGO in person multiple times and attended meetings after initial phone calls. We planned to design our questionnaire on basic metadata about the NGOs, members, and related problems. The interviewees were the administrator or founder of each NGO and four other members in that NGO, in total 5-6 members. The interview duration was about $1.5$ to $2$ hours and we took numerous notes, comments, in some cases recorded the audio conversation with permission. The age range of each interviewee was $20$ to $25$ years. The language of the interview was in Bengali.

At the outset, since the number of volunteers and external funds are most vital for the sustenance of welfare organizations, eventually, our interviews centered on these topics more than others. Firstly, we wanted to find out how a common citizen can know about such organizations if he/she wants to be a part of the organization by becoming a volunteer or by donating funds. The answer is not straightforward. As stated earlier, there is no specific list or database of such organizations, and it is not easy for a common citizen to find their existences. Current volunteers informed us that the effort to join/ contribute to a welfare organization in Bangladesh can be so strenuous that many are dissuaded from even attempting to do so.  This was a serious challenge identified in our discovery process.      

\begin{figure}[h!]
             \centering
            \includegraphics[width = 0.48\textwidth]{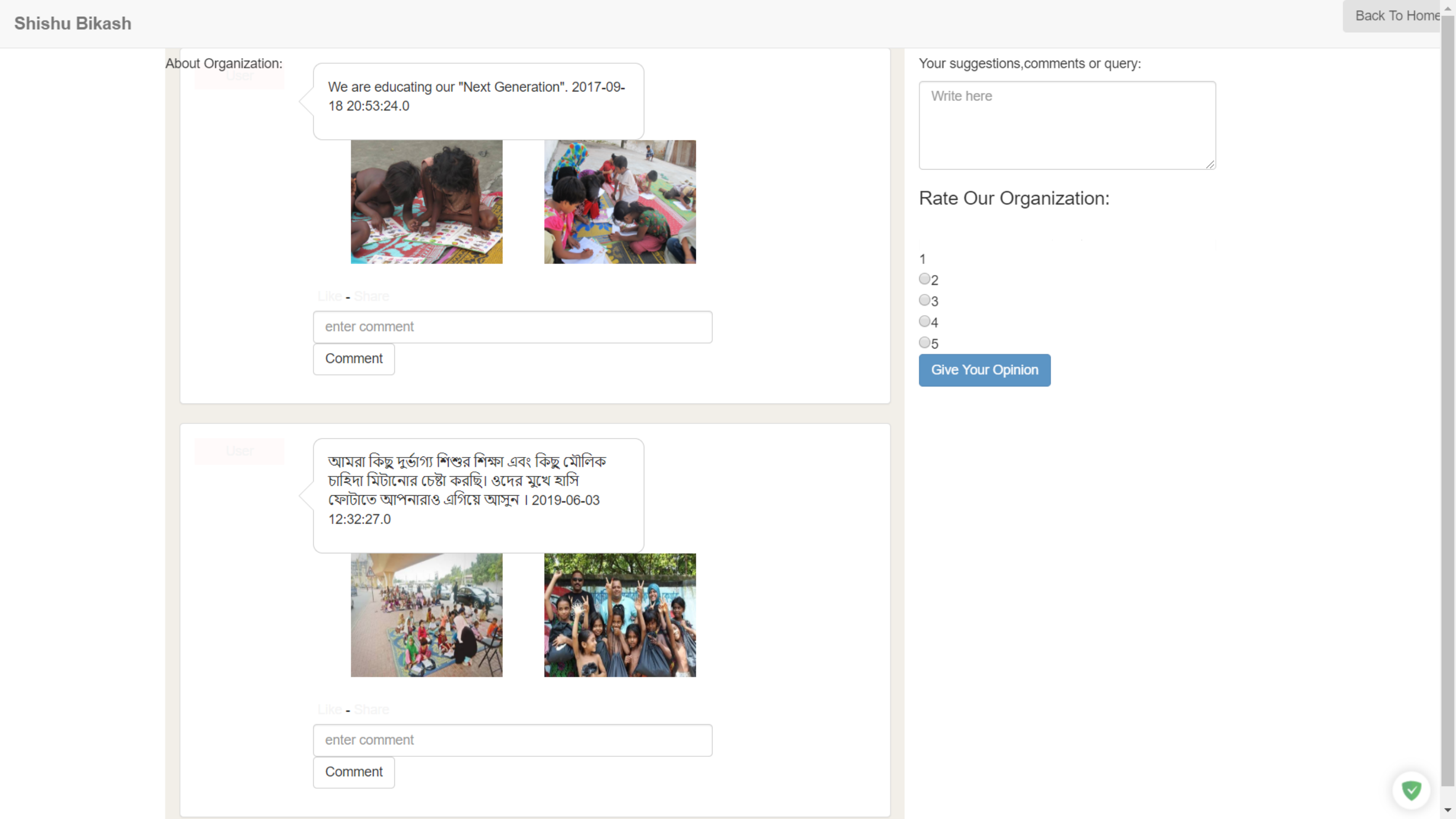}
            \caption{Ability of NGOs to showcase their activities through our system}
            \label{fig:Organization activity showcasing}
             \end{figure}
Secondly, we find that organizations are currently unable to showcase their voluntary activities effectively to the general public. What is sadder is that all interviewees knew the importance of doing so. Here, financial and time constraints are the most notable barriers. Small organizations are forced to use every Taka (Bangladesh Currency) that they receive to satisfy basic needs of street children, and nothing is left for promotional activities. With whatever is left, they end up putting fliers and posters on roads, bus stations, and train stations, and the general public hardly notices them in the realm of numerous posters of other purposes (such as political, job, religious, etc.) aside. The problem is more complicated since more established agencies with their existing clout do attract funds from donors, however, these donors that fund such big agencies are simply unaware of local organizations in their neighborhoods, only because there is no mechanism for them to know. With these barriers, it became clear that nothing else can be done to improve this situation, and volunteers we spoke also agreed to this point. Thirdly, volunteers also mentioned that in low economy countries, donors are fewer in number, and those that are willing to contribute funds expect high standards in accountability and credibility, and demonstrating this to donors was also identified as a major challenge.  

  \begin{figure}[h!]


\includegraphics[width = 0.48\textwidth]{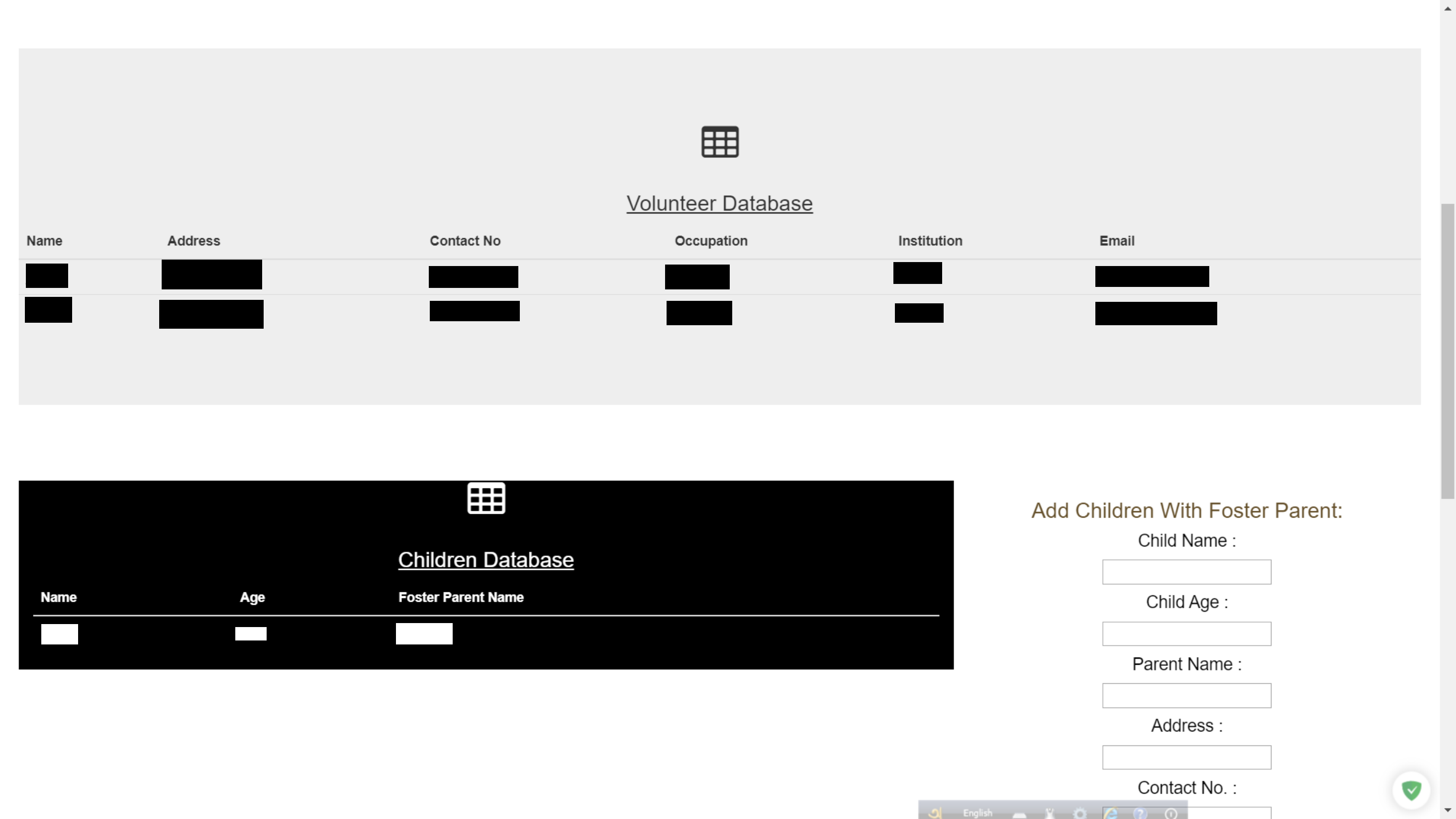}
\caption {Information management system of volunteer and children of BONDHON}
\label{fig:Information management system of volunteer and children of BONDHON}
\end{figure}

\vspace{-1ex}
We now present critical features of our interview process, and how we arrived at the above conclusions. First off, it took us a month to finish all interviews and digest responses. We followed the thematic analysis \cite{thematic} to transcribe our collected information. A total of $15$ volunteers spread evenly across the three NGOs were the interviewees. We repeatedly read all notes collected during interviews, listened to audio recording many times, and discussed findings among ourselves many time overs to arrive at our conclusions. All conclusions identified above were similar for all three NGOs (two micro and one semi-micro). To reiterate, the key findings are-- a) the impact that small scale NGOs created for street children by proving for their most basic needs like food, medicine, clothes, and shelter are vital in their local neighborhoods; b) the NGOs suffer from low membership numbers, and more importantly scant funds, which is unsustainable; c) merely putting up fliers on public spaces does not provide any benefit to these organizations, since potential donors simply do not see these fliers, or are not impressed enough; and d) a digital hub that can create effective engagement between NGOs and general public are critically important for showcasing activities, demonstrating impacts, and improving credibility/ accountability of NGOs, which when done may attract more members and funds that can be used by NGOs to improve lives of street children.

\subsection{Design and Development}

After our first round of interviews, we were able to understand a clear overview of NGOs, demographic information of volunteers, and the problems faced by them while running these organizations. Through iterative processing and reviewing of the data (notes and audio records), we found a common keyword \textbf{Effective Engagement} from all of three NGOs in addition to identifying unique details on NGOs' organizational structure. However, we were also excited in that we were confident now to design a technology-based solution to address their most critical need - how to effectively showcase their activities via designing a deploying a digital hub to connect volunteers, donors, and the general public. Fig. \ref{fig:Block Diagram} presents a simple block diagram of our design in this regard. Based on the design, we proceeded to develop an integrated participatory organization hub encompassing diverse welfare organizations and called our system \textbf{BONDHON} (in Bengali language BONDHON means strong ties with people). In \textbf{BONDHON}, our goal is to bring organizations under a unified platform, while still preserving respective individualities. Most importantly, the general public who are willing to volunteer or donate to smaller NGOs now have \textbf{ONE} unified web platform to engage with members. Volunteers and organizations can use our free portal to enter their credentials, and can immediately be noticed based on featured services provided or locations (as queried by the general public) without a) the organizations having to spend time and money on publicity, and b) the general public struggling to locate organizations in and around where they want to help\footnote{We do not specify the URL of our web portal in this paper to prevent author identification while paper is under review.}.
 
 \begin{figure}[h!]
             \centering
            \includegraphics[width = \linewidth]{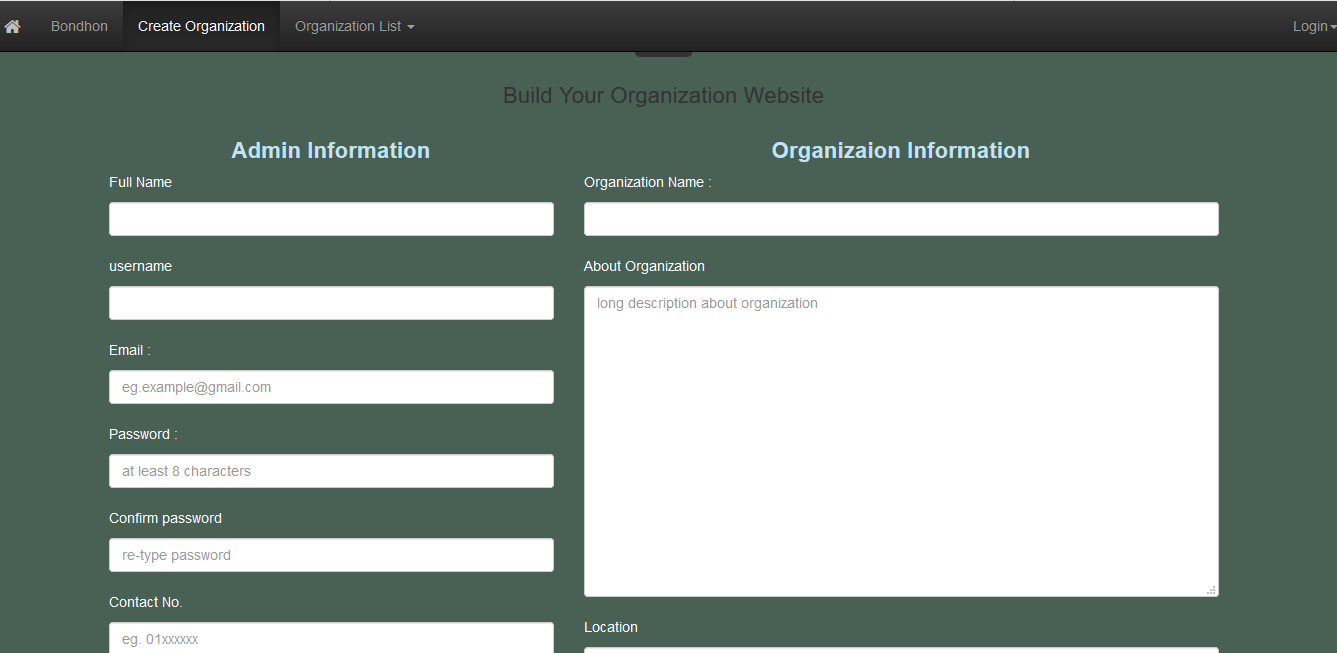}
            \caption{Creating a new organization in BONDHON}
            \label{fig Creating new organization page}
             \end{figure}

Our system was developed in Java using the Spring MVC Framework. The front end was developed using html, css, javascript, bootstrap, and the database was developed using MySql. We have hosted it on Amazon Web Services (AWS). In our web-based system, where a) volunteers in any NGO can manage their data of existing children and services better; create simple dashboards; maintain records of activities; upload videos, pictures, and promotional materials in an engaging form, see outstanding requests for membership or comments or intents for funds from the general public and more; b) prospective volunteers from the general public can search for any organization by interest or zip code or colloquial names and volunteer can join as a member; and c) donors can do the above, browse activities, see  promotional materials, and  have a mechanism to express intents to donate funds.


In a nutshell, our initial prototype offers the following services based on stakeholders:
        \begin{itemize}
         \item \textbf{Virtual Organization:} A publicly accessible platform that enables members in any existing NGO to create a specific portal for that particular organization which is shown in Fig. \ref{fig Creating new organization page}. Databases pertaining to the organization can be custom created and maintained. Outstanding requests for membership or questions or comments can be viewed and responded to. The portal easy uploads of photos, videos and other forms of promotional materials with keywords, which are searchable and viewable. Fig. \ref{fig:Organization activity showcasing} shows some representative images that could be uploaded. We were motivated from the following speech regarding the feature-- \\
         
         \textit{We cannot publicize our daily activities like schooling and other events like free health campaign at a large scale. We do need a way to showcase our activities to a wider scale and obviously at a low cost. Postering, flyers are cost centers for us right now.}
         
        \item \textbf{Smart Information Management:}  Any person willing to provide voluntary services at any NGO or who wants to become a member can apply for membership at that specific organization after filling out information on demographics, interests, and experiences. The information is automatically sent to members who could approve, and the response is conveyed via an email back to the prospective volunteer. Administrators can control the volunteer and children management system with necessary information as they see fit as shown in Fig. \ref{fig:Information management system of volunteer and children of BONDHON}.
        
        \item \textbf{Foster Parent Management} Children being serviced by an organization are now cataloged. In Bangladesh, children can be adopted by foster parents, however, such children need a different kind of tracking mechanism since they now reside in another home. In our portal, separately managing foster children from others is enabled as seen in Fig.  \ref{fig:Information management system of volunteer and children of BONDHON}. One owner of the organization stated during the interview--\\
        
        \textit{Since these children are vulnerable and can go anywhere at any time, we try to engage foster parent to take the responsibility of at least one children. It is obvious that we cannot manage the whole process efficiently in many cases.}

        \item  Any citizen can view all the activities of any NGO without any restrictions/ obligations. Moreover, any individual can express his/her opinion through comment(s) on any particular post of an organization as well as can rate any organization on a scale of $1$ to $5$ (Likert scale) based on its activities. Thus, the members of an organization can get feedback from the general public, which is critical for their continued sustenance.
    
        \item The general public can visit this portal to search for NGOs using zip code, common colloquial names, kinds of activities performed (e.g., education vs. healthcare vs. adoption, etc.), and then organizations' matching criteria are easily retrieved. Our portal also enables donors to contribute funds to organizations. We set a limit of  $10,000$ BDT (around $125$), as per  legal restrictions. A very popular, quick and easy-to-use mobile transaction environment called  `bKash'\footnote{\url{https://www.bkash.com/}} is used for the money transaction. It is an established system in Bangladesh for small money transaction. It requires just a cell phone for the recipient and hence making it easy to use. Depending on the success of our current system, a payment gateway, if asked from the stakeholders, for larger transactions will be set-up soon.
\end{itemize}

To summarize here, we have voluntarily created a web-portal with the theme of \textbf{Effective Engagement} so that the general public, donors, and NGOs can stay engaged with each other. Our design attempted to integrate critical requirements of three NGOs we surveyed, which is to a) enable increased publicity for NGOs; b) enable volunteers to join NGOs as members; and c) create opportunities for donors to locate, engage, and reform possible fund-related activities. Next, we present the results of a phase of our deployment.

\begin{figure}[htbp]

\minipage[t]{0.48\linewidth}
\includegraphics[width=\linewidth, height = 3cm ]{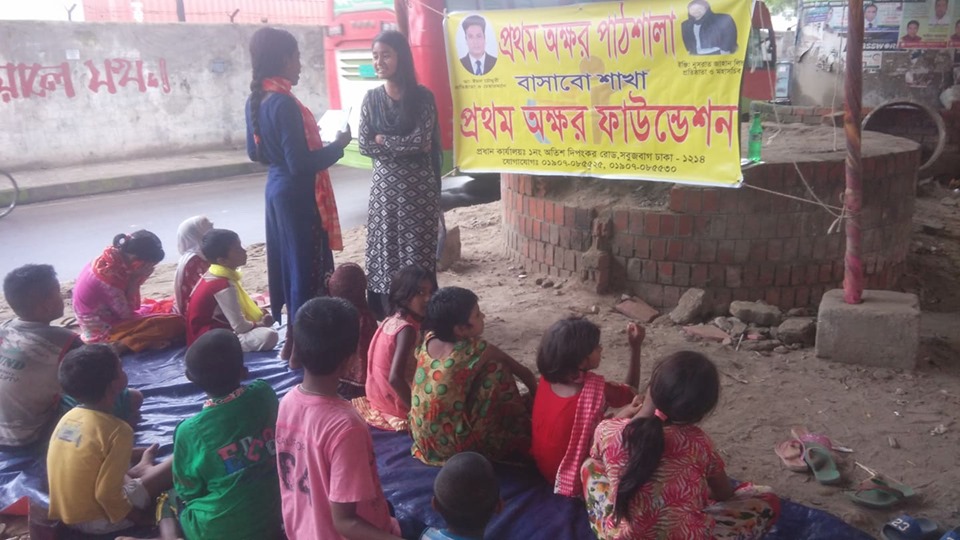}
  \subcaption{A schooling session arranged by an organization named ``PROTHOM AKKHOR FOUNDATION''}\label{fig:A schooling session arranged by an organization prothom akkhor founadtion}
\endminipage
\hfill
\minipage[t]{0.48\linewidth}
  \includegraphics[width=\linewidth, height = 3cm]{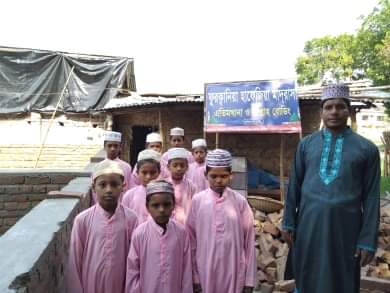}
  \subcaption{A campaign arranged by an organization named ``Siddiqia Bissho Islami Mission''}\label{fig:A schooling session arranged by an orphanage}  
\endminipage\hfill

\caption {Campaign and schooling arranged by an NGO and an orphanage}
\label{fig:Schooling and campaign arranged by different orphanage and NGOs}

\end{figure}

\subsection{Deployment, Usage, and Responses}
 
It took us eight weeks to develop the platform and test it to our satisfaction. Then, we contacted members of the three NGOs whom we met earlier as well as two new organizations named {\em PROTHOM AKKHOR FORUNDATION} and {\em Siddiqia Bissho Islami Mission} (an orphanage for Muslim children) to request them to evaluate our system. Recall that since these are young people with university degrees (or in the process of a degree) in case of {\em PROTHOM AKKHOR FORUNDATION} similar to our earlier organizations, they were well versed in digital/ web technologies, and how to evaluate them. Besides, the engagement of {\em Siddiqia Bissho Islami Mission} was done through one of the authors on behalf of the organization, as personnel of this organization is not that experienced from the technological perspectives shown in Fig. \ref{fig:Schooling and campaign arranged by different orphanage and NGOs}.

The motive behind bringing {\em Siddiqia Bissho Islami Mission} in our system is to explore how people react with an orphanage beside organizations for street children. Besides, the orphanages in Bangladesh generally receive some donations from general people. Therefore, we intended to explore whether having an orphanage in our system could enable making such donations. Evaluation of inclusion of this orphanage can be done through such donations if any. We will present the extent of the donations and their nature later in the paper. This orphanage has around 30 children and fully rely on the mass people donations. This type of organizations is also operated by a founding body. The second new organization has 20 volunteers, the primary functionality is to run school among the street children of the southern part of Dhaka city. The organization has around 50 children in their school and donation is collected from the volunteers and internal donors.

Besides, all members in {\em Child Care}, and {\em Putul}, {\em PROTHOM AKKHOR FORUNDATION}, were also eager to evaluate our system, and it was very easy to convince them once they saw a one-hour demonstration. On the other hand, it was not easy convincing {\em Shishu Bikash} to use our system, for reasons we will explain later. As such, members of organizations {\em Child Care} and {\em Putul} tested our system for a period of $30$ days.  We set up Google Analytics into our system to assess their engagement. For example, in  October $2018$, the total number of users was $73$ ($63$ new users and $10$ returning users). The total count of \textit{pages viewed} was $1605$, the average number of pages viewed per session was $6.30$;  average session duration was $7$ min and $38$ sec, and the bounce rate was $11.36\%$.

\vspace{-1ex}

\subsection{Findings and Insights from Second Face-to-Face Interviews}

After the $30$--day evaluation of our system, we scheduled a second round of semi-structured face-to-face interviews with admins, founders, and volunteers of the previous three organizations along with the newer two NGOs that participated in the evaluation. Six participants from one organization and eight from each from the other four participated in one on one interviews. They included the founder/ admin of each organization. This time we set $20$ questions to gauge their feedback on our system for its evaluation and we also see if we rightly got their problems right after the earlier interviews. Each interview lasts for around two hours. Some specifics whether we were interested in their suggestions on the user interface, ease of overall use, whether their core requirements were satisfied, and whether or not they agreed that our portal eases their pains towards external engagement. The entire duration to complete all interviews was a month. Numerous notes, records, and audio clips were collected and analyzed with thematic process by our team after the interview.

\begin{figure}[h!]
         \centering
        
        \includegraphics[width = \linewidth]{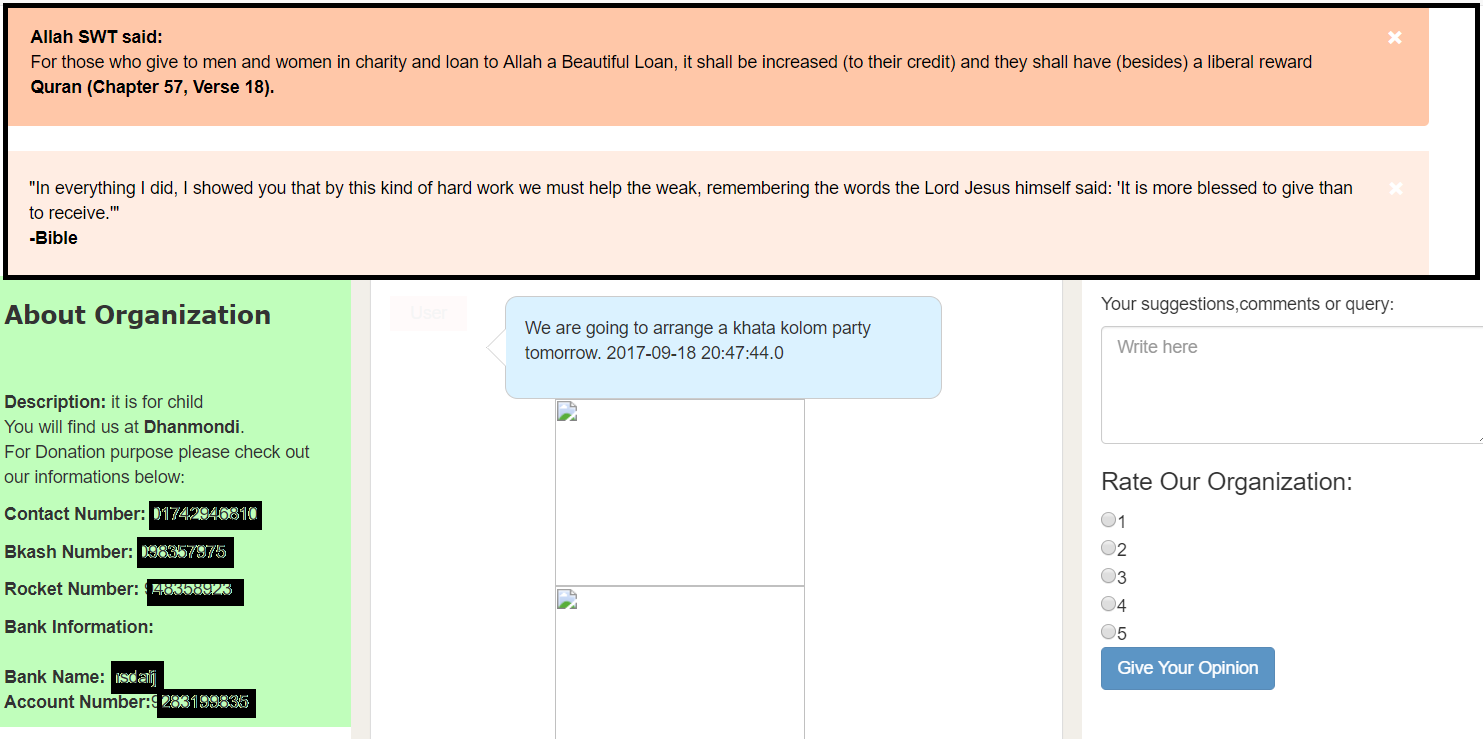}
         
        \caption{Religious quote enclosed in black rectangle to encourage moral norms}
        \label{fig:Religious quote enclosed in black rectangle, rating options }
     \end{figure}

At the outset, using the thematic analysis for transcribing the notes, all interviewees agreed that our proposed system can be very important to augment their efforts. There were only minor comments about the user interface, our dashboards design, etc. All interviewees agreed that the open nature of our system meant that members of NGOs can now freely showcase their activities to the outside world. They all agreed that this approach to enhance engagement was vital for them to both increase membership, and also attract external funds more confidently, and with more credibility.

One concern they raised was preserving the core individualities of their organizations. Our discussions with them revealed that volunteers have their personal views on our system that may conflict with views of other members of the same NGO. The diversity is more prominent in the semi-micro organization, despite the common theme of working for street children. Some differences arose in regarding how to handle events of organizations, how to present policies to the public, how to manage donors, how to give feedback to donors, and how to enhance their campaigns with our system in place. We finally concluded that our system while being stable and useful needs a ground-up refinement following the Value Sensitive Design (VSD) \cite{VSD} for wide-scale usability as well as acceptability. We are now doing this by passively observing NGO meetings, health/ educational camps, campaign activities of NGOs, donor meetings, and much more to document related activities and re-design our system with VSD principles. Next, we present findings for future guidance in this aspect.

\subsection{Values in Design}

The stakeholders of our system, i.e., NGOs and donors hold a common value towards upgrading lifestyle for the street children. To address the values through the principle of the VSD, we have explored different aspects presented in the earlier section. The challenging part is here how a system can realize all the values effectively. 

In case of the Western culture, the \textit{Theory of Planned Behavior} (TPB) \cite{TPB} model has been employed to find the intention of charity donations \cite{knowles2012}.   However, in the case of developing countries, several variables are used for understanding charity donations \cite{rogers2018social}. Firstly, subjective norms and attitude are recognized as strong predictors of donation behaviour \cite{fishbein1977}. Accordingly, we put the donors recognition (by taking the permission from donors) in front of public, they should be get encouraged to donate again. From the perspective of VSD, we have made the donation information publicly available for respective organizations. Secondly, for injunctive and descriptive norms \cite{warburton2000volunteer, mcmillan2003using},  we can disclose identity of each commentators and donors so that if someone see people in his group to make donations he may also come toward to donate. Based on this, we go for a ``I vote" button on each post to make the donors interested. Thirdly, aside from this, moral norms have also been used to predict altruistic behaviour in studies pertaining to money donation \cite{warburton2000volunteer,armitage2001efficacy}. To encourage such moral norms we have added some religious quotes in our platform of each organization shown in Fig. \ref{fig:Religious quote enclosed in black rectangle, rating options }

 \vspace*{-5px}



\subsection{Changing the Scenario} During the second interview, we have interviewed two new organizations (``PROTHOM AKKHOR FOUNDATION" and ``Siddiqua Bissho Islami Mission"). Among these two organizations, ``PROTHOM AKKHOR FOUNDATION" represents an organization of street children. Besides, ``Siddiqua Bissho Islami Mission" represents an orphanage. Note that an orphanage has some baseline similarities with an organization for street children, as both the types of organizations are a charity in nature having wretched poverty-stricken children. Here, a difference between the two types of organizations lies in the fact that the orphanage hosts its service recipients, i.e., the orphans, whereas the organization for street children does not do so.

These two new organizations act as validators of our design that was an outcome of the feedback from the earlier organizations. These new organizations also agreed with the design and we have found a notable donor--organization interactions that were hardly seen previously (before the iteration of VSD). After adopting the system, in the case of the ``Siddiqua Bissho Islami Mission", we have received around 107k BDT (1278 USD) as donation within a month after introducing it in our system. Here, among the six donors, male and female percentages of donors are 83.33\% and 16.67\% respectively. Most importantly, after the iteration of VSD, we can attract new donors for this organization. This happening of attracting new donors indicate the success of iteration of VSD principle into the design, which eventually resulted in an enhanced donation. Here, it is noteworthy that we introduced ``Siddiqua Bissho Islami Mission" before the Arabic month of Ramadan, which is treated as a holy month for the Muslims (the majority in Bangladesh). Most (four out of six) of the donations we received for the organization was during the month of Ramadan. Notably, all the donations were made by Muslims. This happening suggests that there exist some triggering with the month having religious value. We elaborate this happening more later in this paper in the context of HCI.

 \section{Confusions and Questions about Resources, Competition, and Donors}
 \label{sec: Confusions}

We now present additional insights into impending complexities as we move forward with our system design. For example, one organization requested an option for advisory panel management, which others did not ask for. We have incorporated it for them. However, in the case of semi-macro organizations, the situation is a bit more complex, where organizational decisions are taken after discussions among volunteers, founders, and advisory panels. Informal communications between our team and other semi-micro organizations revealed that some semi-micro NGOs are willing to accept our system as a showcasing platform after a lot of discussions and arguments. Some other NGOs are still in a fix as to whether or not they should be using our system.

Careful discussions revealed that as NGOs grow more in member size, donor funding is paramount for success. This makes sense since large organizations cannot succeed without external funds. The semi-micro organizations are worried that an integrated platform that will be shared by multiple NGOs may decrease their chances of securing funding. One founding member of the semi-micro organization {\em Shishu Bikash} mentioned this point as follow

 \textit{``We appreciate the idea of organization hub, however, our organizations are yet to take some time to take decisions since our acting body are divided among them about the decision whether we should join or not. We are trying to build our website rather. We are not sure about how people will get us and how the engagement can enhance our activities, and most importantly how we can reach more donors. If there are many organizations in your hub, then there is a  sharp chance of  losing our donors."}
 
It is not too hard to understand this sentiment. In poorer countries, there will be intense competition for donor support, and an avenue that can potentially decrease donor support will be viewed very negatively by NGO members. This is especially true for larger-scale organizations where decisions take time. Hence, a need to maintain status-quo may be seen as a good option sometimes by members.  We present some perspectives on this now. Our discussions revealed that funding for smaller--scale and independently-run NGOs are geographic specific. Hence, we provide location-based services in our system wherein NGOs can identify areas they operate, and donors can choose to view NGOs based on location. We expect this to alleviate some concerns regarding competition among resources. Furthermore, most NGO members do agree that the ultimate goal is to improve the lives of street children, and therefore, any hub that will enable organizations to see and share what others are doing will be beneficial overall to learn best practices, which our system enables seamlessly.

The question regarding the loss of potential donors needs some explanation though. In the case of making donations, citizens often have a concern for credibility and accountability. In this context, any platform that enables effective advertisement is vital to seek funding. Donors also want to see how their money is being used for donating further, and digital platforms are most suitable to satisfy this expectation. Moreover, we believe that competition breeds creativity, and NGOs that do better will be rewarded better. Furthermore, different organizations can have different focus points (e.g., education vs. healthcare vs. medicines vs. enabling adoption, etc.), therefore, when multiple organizations can be showcased at the same time, different donors may see things differently and spread their donations among NGOs as they see fit. This may be a welcome motivation for NGOs to engage with our platform. One donor's comment summarizes our perspective best. The donor says that--

\textit{``There are many people who want to donate, but the fact is we don't know where to donate and you know what, the most important thing is trust. I want to know how my money is being utilized and I want accountability and credibility. Again, people want trustworthy organizations and the more volunteers work, the more they can be trusted. A public open platform can gear up organizations to do better work than another and certainly, competition for good work is not bad at all."}

Our platform currently provides a platform for donors to engage with NGO members. Donors can ask questions, see responses, provide comments, and also like/ dislike activities. In this manner, we believe that many concerns regarding donor support to NGOs are mitigated, and resulting in long-lasting, meaningful, and credible interactions. While we are excited at these developments, still some challenges remain, which we present next.

\section{Impacts of Human/ Organizational Factors on Our System}

\label{sec:Human factor}
After the deployments of the system among the organizations, we faced human and organizational factors that limit the scope of the deployment in semi-micro organizations. These are stated below--

\subsection{Mistrust Among NGOs in a Shared Platform}

In some sense, founders of small--scale NGOs working for street children can be called ``social entrepreneurs", who work to collect and maximize funds to develop, fund, and implement solutions to social, cultural, or environmental issues. As such, marketing, publicity, credibility, trust, accountability, legal aspects, unified vision and several more are vital for sustenance of these NGOs. 

We found that despite having common (and noble) goals, there is an inherent sense of fear and mistrust between organizations, due to which NGOs (especially larger ones) are reluctant to participate in shared efforts. This was disappointing for us, however nevertheless understandable, since in poorer countries, any small donation does not come without significant effort. Therefore, in parallel to working for the cause of street children, NGO members need to devote time, energy and sometimes even money to protect and preserve what they are creating/ operating. While small scale organizations are more willing to participate in our platform for engagement, we could sense that while the ability to help street children was their primary agenda, we get a sense that members here are equally excited about growth and publicity of their organization, which will help them attract more funds. Moreover, we are wondering whether or not these micro organizations will still remain engaged with our system should they reach a larger membership number, and hence increase complexities in decision making. This is an open issue for us now.

\section{Discussion}
\label{sec: disc}

After the mixed experience of both success and failure in two different contexts, we look for determining the internal and external variables for the success and failure by conceptualizing the donors and NGOs behavior. During the month of Ramadan, the orphanage achieves a significant amount of donations. In this study \cite{ranganathan2008determinants}, a path model is defined to identify the intention of donor behaviors. The variables of the model are {\em religiosity}, { \em attitude towards helping others (AHO)}, {\em attitude towards charitable organizations (ACO)}, {\em attitude towards the advertisement (Attad)}, and {\em behavioral intentions (BI)} where {\em religiosity} have a direct positive impact on all other factors. The generality of the system needs to be explored by considering the social, cultural, religious context of the other countries. About 99\% of people in Bangladesh are religious minded, therefore, we have tried to conceptualize the norms. Since Bangladesh is a Muslim majority country, during the holy month of Ramadan, the donation rate is high. To this context, our system aligns with the key features (accessibility, accountability, and interaction) of trustworthy web platform for charity donors described here \cite{sargeant, donortrust}. Therefore, micro organizations took the platform as a medium of their publicity towards a wider donor community. However, before a particular time (in the month of Ramadan), the enlisted micro organizations gain a little attraction from the donors and at the month of Ramadan, the donation enhances significantly. Most importantly, the latest organization, which was not involved in our first interview, achieves the success and indicates the validity of the design principles followed in our system since we did not take any explicit requirements from them rather just they just adopt the system. On the other hand, in case of complex organizations, organizational ergonomics, and trust issues still limit the opportunity for semi-micro organizations though they were one of the groups that have participated in all interviews, designs, and feedback of the design. The initial success of the micro organizations might give them a room of reviving trust. Therefore, we find that \textit{time} is an important factor in the case of micro organizations for describing donors behavior in developing country context. In addition, complex \textit{organizational ergonomics}, and \textit{trust} issues should be taken into account.

However, in a country where religious practices hardly affect the social, cultural and behavioral norms, the donor behavior might be different in this context. Besides, the organizational and human factors are studied in the developing countries context. For example, if the platform needs to be deployed in any western countries as a solution of homeless people, the VSD principles of design, as well as donor behavior, needs to be revisited.

\section{Conclusion and Future Work}

\label{sec:Discussion}

Street Children are one of the most vulnerable communities in the world specially in poorer countries. While local (called NGOs) and global organizations are attempting to address this problem in a country like Bangladesh, smaller organizations are disproportionately disadvantaged due to lack of brand name/ publicity. Creating technology-based solutions to cater to smaller-scaled organizations working for the upliftment of street children is hence a critical need of the hour in poorer countries. In doing so, NGOs can be classified into two types - micro and semi-micro according to the behavior of the organization. Considering such a classification, in this study, we have taken two rigorous face to face interviews to identify the actual scenarios and according to the responses we have designed an integrated hub intended for the NGOs. 

During our deployment for the process of evaluation of our system, we have also incorporated two new organizations (one of them is an orphanage) to let the system presented to more general behavior. After the successful donation through the system in micro-organizations, we establish that, we find the positive co relation of donors behavior with relevant study and discussed how the organizational and human factors limit the tendency.


In the future, we plan to expand the scale of our deployment by incorporating more organizations in Bangladesh. Nonetheless, we intend to enlarge it across the border by taking organizations of other similar countries on board. Finally, leveraging our digital hub, we plan to explore the possibility of integrating interactions of the children (served by the organizations) in response to various activities performed by the organizations. Such integration could bring the voice of the service recipients on the surface eventually enabling a 360-degree interaction overall stakeholders of the ecosystem.

\balance{}

\bibliographystyle{SIGCHI-Reference-Format}
\bibliography{sample}


\begin{thebibliography}{00}


\ifx \showCODEN    \undefined \def \showCODEN     #1{\unskip}     \fi
\ifx \showDOI      \undefined \def \showDOI       #1{{\tt DOI:}\penalty0{#1}\ }
  \fi
\ifx \showISBNx    \undefined \def \showISBNx     #1{\unskip}     \fi
\ifx \showISBNxiii \undefined \def \showISBNxiii  #1{\unskip}     \fi
\ifx \showISSN     \undefined \def \showISSN      #1{\unskip}     \fi
\ifx \showLCCN     \undefined \def \showLCCN      #1{\unskip}     \fi
\ifx \shownote     \undefined \def \shownote      #1{#1}          \fi
\ifx \showarticletitle \undefined \def \showarticletitle #1{#1}   \fi
\ifx \showURL      \undefined \def \showURL       #1{#1}          \fi

\bibitem{Street}
 2019.
\newblock Street Children in {Bangladesh}: {A} {Life} {of} {Uncertainity}.
\newblock \url{ http://www.theindependentbd.com/printversion/details/32932}.
  (2019).
\newblock
\newblock
\shownote{Accessed: September 15, 2019.}


\bibitem{TPB}
{Icek Ajzen}. 1991.
\newblock \showarticletitle{The theory of planned behavior}.
\newblock {\em Organizational behavior and human decision processes\/} {50}, 2
  (1991), 179--211.
\newblock


\bibitem{armitage2001efficacy}
{Christopher~J Armitage} {and} {Mark Conner}. 2001.
\newblock \showarticletitle{Efficacy of the theory of planned behaviour: A
  meta-analytic review}.
\newblock {\em British journal of social psychology\/} {40}, 4 (2001),
  471--499.
\newblock


\bibitem{thematic}
{Jodi Aronson}. 1995.
\newblock \showarticletitle{A pragmatic view of thematic analysis}.
\newblock {\em The qualitative report\/} {2}, 1 (1995), 1--3.
\newblock


\bibitem{DIKC}
{Carlos Bilich}. 2007.
\newblock \showarticletitle{How can ICT be Used to Improve Street Children's
  Plight: A Proposal for the City of Santa Fe, Argentina}.
\newblock  (2007).
\newblock


\bibitem{brown1991bridging}
{L~David Brown}. 1991.
\newblock \showarticletitle{Bridging organizations and sustainable
  development}.
\newblock {\em Human relations\/} {44}, 8 (1991), 807--831.
\newblock


\bibitem{donortrust}
{Christopher~DB Burt}. 2014.
\newblock {\em Managing the Public's Trust in Non-profit Organizations}.
\newblock Springer.
\newblock


\bibitem{effectiveComputing}
{M.~Bernardine Dias} {and} {Eric Brewer}. 2009.
\newblock \showarticletitle{How Computer Science Serves the Developing World}.
\newblock {\em Commun. ACM\/} {52}, 6 (June 2009), 74--80.
\newblock
\showISSN{0001-0782}
\showDOI{%
\url{http://dx.doi.org/10.1145/1516046.1516064}}


\bibitem{fishbein1977}
{Martin Fishbein} {and} {Icek Ajzen}. 1977.
\newblock \showarticletitle{Belief, attitude, intention, and behavior: An
  introduction to theory and research}.
\newblock  (1977).
\newblock


\bibitem{VSD}
{Batya Friedman}. 1996.
\newblock \showarticletitle{Value sensitive design}.
\newblock {\em interactions\/} {3}, 6 (1996), 16--23.
\newblock


\bibitem{goecks2008charitable}
{Jeremy Goecks}, {Amy Voida}, {Stephen Voida}, {and} {Elizabeth~D Mynatt}.
  2008.
\newblock \showarticletitle{Charitable technologies: Opportunities for
  collaborative computing in nonprofit fundraising}. In {\em Proceedings of the
  2008 ACM conference on Computer supported cooperative work}. ACM, 689--698.
\newblock


\bibitem{knowles2012}
{Simon~R Knowles}, {Melissa~K Hyde}, {and} {Katherine~M White}. 2012.
\newblock \showarticletitle{Predictors of young people's charitable intentions
  to donate money: An extended theory of planned behavior perspective}.
\newblock {\em Journal of Applied Social Psychology\/} {42}, 9 (2012),
  2096--2110.
\newblock


\bibitem{perception}
{Christopher~A. Le~Dantec} {and} {W.~Keith Edwards}. 2008a.
\newblock \showarticletitle{Designs on Dignity: Perceptions of Technology Among
  the Homeless}. In {\em Proceedings of the SIGCHI Conference on Human Factors
  in Computing Systems} {\em (CHI '08)}. ACM, New York, NY, USA, 627--636.
\newblock
\showISBNx{978-1-60558-011-1}
\showDOI{%
\url{http://dx.doi.org/10.1145/1357054.1357155}}


\bibitem{Organization}
{Christopher~A. Le~Dantec} {and} {W.~Keith Edwards}. 2008b.
\newblock \showarticletitle{The View from the Trenches: Organization, Power,
  and Technology at Two Nonprofit Homeless Outreach Centers}. In {\em
  Proceedings of the 2008 ACM Conference on Computer Supported Cooperative
  Work} {\em (CSCW '08)}. ACM, New York, NY, USA, 589--598.
\newblock
\showISBNx{978-1-60558-007-4}
\showDOI{%
\url{http://dx.doi.org/10.1145/1460563.1460656}}


\bibitem{mcmillan2003using}
{Brian McMillan} {and} {Mark Conner}. 2003.
\newblock \showarticletitle{Using the theory of planned behaviour to understand
  alcohol and tobacco use in students}.
\newblock {\em Psychology, Health \& Medicine\/} {8}, 3 (2003), 317--328.
\newblock


\bibitem{merkel2007managing}
{Cecelia Merkel}, {Umer Farooq}, {Lu Xiao}, {Craig Ganoe}, {Mary~Beth Rosson},
  {and} {John~M Carroll}. 2007.
\newblock \showarticletitle{Managing technology use and learning in nonprofit
  community organizations: methodological challenges and opportunities}. In
  {\em Proceedings of the 2007 symposium on Computer human interaction for the
  management of information technology}. ACM, 8.
\newblock


\bibitem{sustainable}
{B{\^a}c~Dorin Paul}. 2008.
\newblock \showarticletitle{A history of the concept of sustainable
  development: Literature review}.
\newblock {\em The Annals of the University of Oradea\/} {17}, 2 (2008), 581.
\newblock


\bibitem{population}
{Bangladesh {Population}}. 2019.
\newblock   (2019).
\newblock
\showURL{%
\url{http://worldpopulationreview.com/countries/bangladesh-population/}}
\newblock
\shownote{Accessed: September 15, 2019.}


\bibitem{interactivetool}
{H~Vignesh Ramamoorthy}, {PJ Balakumaran}, {and} {H Karthikeyani}. 2013.
\newblock \showarticletitle{An Interactive Teaching-Learning Tool for
  Underprivileged Children in Rural Schools}.
\newblock {\em International Journal of Modern Education and Computer
  Science\/} {5}, 7 (2013), 27.
\newblock


\bibitem{ranganathan2008determinants}
{Sampath~Kumar Ranganathan} {and} {Walter~H Henley}. 2008.
\newblock \showarticletitle{Determinants of charitable donation intentions: a
  structural equation model}.
\newblock {\em International Journal of Nonprofit and Voluntary Sector
  Marketing\/} {13}, 1 (2008), 1--11.
\newblock


\bibitem{homelessness}
{Jahmeilah Roberson} {and} {Bonnie Nardi}. 2010.
\newblock \showarticletitle{Survival Needs and Social Inclusion: Technology Use
  Among the Homeless}. In {\em Proceedings of the 2010 ACM Conference on
  Computer Supported Cooperative Work} {\em (CSCW '10)}. ACM, New York, NY,
  USA, 445--448.
\newblock
\showISBNx{978-1-60558-795-0}
\showDOI{%
\url{http://dx.doi.org/10.1145/1718918.1718993}}


\bibitem{rogers2018social}
{Todd Rogers}, {Noah~J Goldstein}, {and} {Craig~R Fox}. 2018.
\newblock \showarticletitle{Social mobilization}.
\newblock {\em Annual review of psychology\/}  {69} (2018), 357--381.
\newblock


\bibitem{humanfactors}
{Gavriel Salvendy}. 2012.
\newblock {\em Handbook of human factors and ergonomics}.
\newblock John Wiley \& Sons.
\newblock


\bibitem{sargeant}
{Adrian Sargeant}, {John~B Ford}, {and} {Jane Hudson}. 2008.
\newblock \showarticletitle{Charity brand personality: The relationship with
  giving behavior}.
\newblock {\em Nonprofit and Voluntary Sector Quarterly\/} {37}, 3 (2008),
  468--491.
\newblock


\bibitem{slay2006bingbee}
{Hannah Slay}, {Peter Wentworth}, {and} {Jonathon Locke}. 2006.
\newblock \showarticletitle{BingBee, an information kiosk for social enablement
  in marginalized communities}. In {\em Proceedings of the 2006 annual research
  conference of the South African institute of computer scientists and
  information technologists on IT research in developing countries}. South
  African Institute for Computer Scientists and Information Technologists,
  107--116.
\newblock


\bibitem{combodia}
{Lindsay Stark}, {Beth~L. Rubenstein}, {Kimchoeun Pak}, {Rosemary Taing}, {Gary
  Yu}, {Sok Kosal}, {and} {Leslie Roberts}. 2017.
\newblock \showarticletitle{Estimating the size of the homeless adolescent
  population across seven cities in Cambodia}.
\newblock {\em BMC Medical Research Methodology\/} {17}, 1 (26 Jan 2017), 13.
\newblock
\showISSN{1471-2288}
\showDOI{%
\url{http://dx.doi.org/10.1186/s12874-017-0293-9}}


\bibitem{palestine}
{Oliver Stickel}, {Dominik Hornung}, {Konstantin Aal}, {Markus Rohde}, {and}
  {Volker Wulf}. 2015.
\newblock \showarticletitle{3D Printing with marginalized children—an
  exploration in a Palestinian refugee camp}. In {\em ECSCW 2015: Proceedings
  of the 14th European Conference on Computer Supported Cooperative Work, 19-23
  September 2015, Oslo, Norway}. Springer, 83--102.
\newblock


\bibitem{ict4d}
{Tim Unwin}. 2009.
\newblock {\em ICT4D: Information and communication technology for
  development}.
\newblock Cambridge University Press.
\newblock


\bibitem{warburton2000volunteer}
{Jeni Warburton} {and} {Deborah~J Terry}. 2000.
\newblock \showarticletitle{Volunteer decision making by older people: A test
  of a revised theory of planned behavior}.
\newblock {\em Basic and Applied Social Psychology\/} {22}, 3 (2000), 245--257.
\newblock


\bibitem{comenet}
{Volker Wulf}, {Volkmar Pipek}, {David Randall}, {Markus Rohde}, {Kjeld
  Schmidt}, {and} {Gunnar Stevens}. 2018.
\newblock {\em Socio-Informatics: A Practice-Based Perspective on the Design
  and Use of IT Artifacts}.
\newblock Oxford University Press.
\newblock


\end{thebibliography}

\end{document}